\documentclass[aps, prd, amsmath, floats, floatfix, twocolumn, nofootinbib, superscriptaddress]{revtex4-1}
\usepackage[pdftex]{graphicx}
\usepackage{color}
\usepackage{latexsym}
\usepackage[colorlinks]{hyperref}
\usepackage{amsmath}

\newcommand{\be}{\begin{eqnarray}}
\newcommand{\ee}{\end{eqnarray}}

\begin{document}

\title{Astrophysical Wormholes}

\author{Cosimo~Bambi}
\email[]{bambi@fudan.edu.cn}
\affiliation{Center for Field Theory and Particle Physics and Department of Physics, Fudan University, 200438 Shanghai, China}

\author{Dejan~Stojkovic}
\email[]{ds77@buffalo.edu}
\affiliation{Department of Physics, State University of New York (SUNY) at Buffalo, Buffalo, NY 14260-1500, United States}

\begin{abstract}
Wormholes are hypothetical topologically-non-trivial structures of the spacetime. From the theoretical point of view, the possibility of their existence is challenging but cannot be ruled out. This article is a compact and non-exhaustive review of past and current efforts to search for astrophysical wormholes in the Universe.
\end{abstract}

\maketitle

\section{Introduction}

\begin{figure}[b]
\centering
\includegraphics[width=8.5cm]{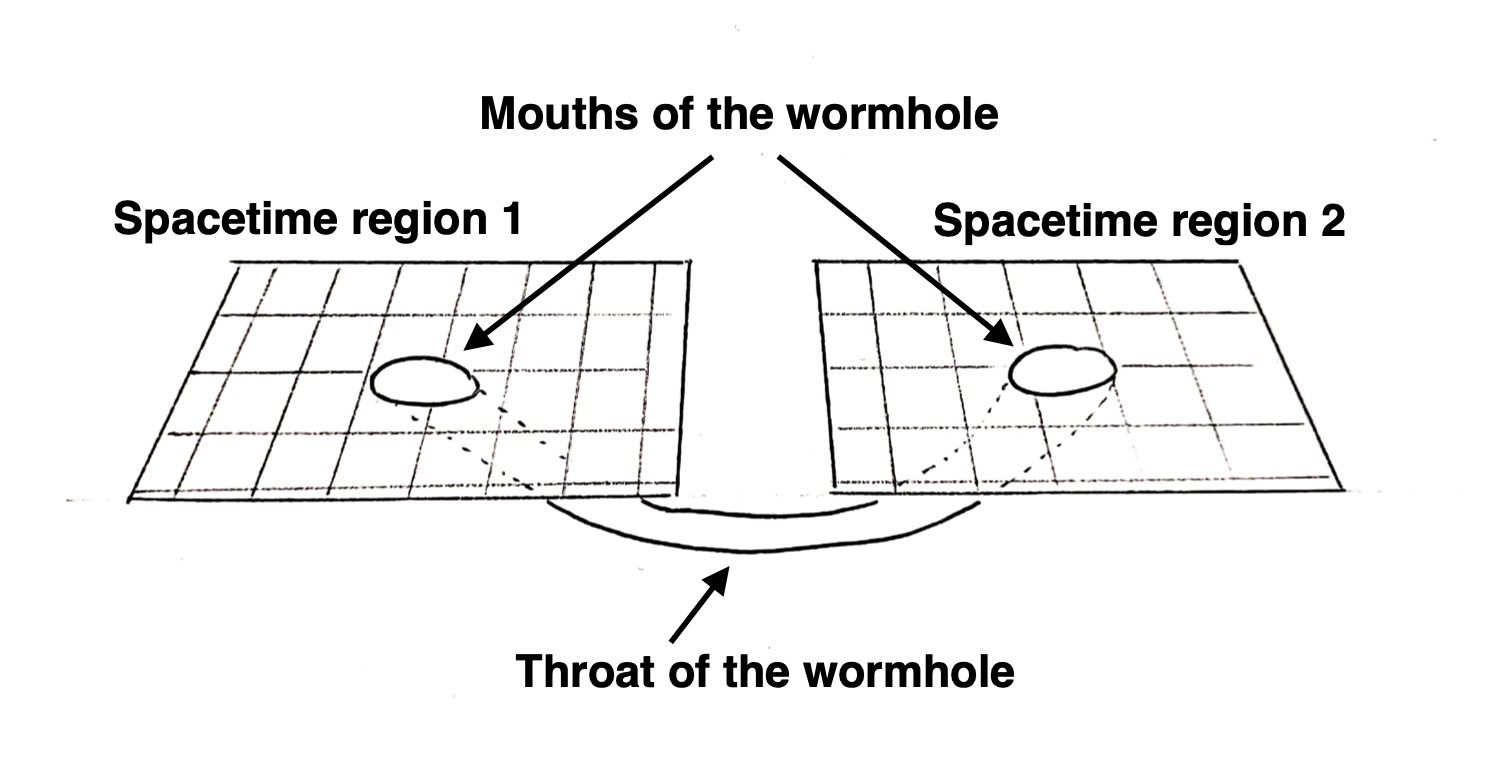}
\vspace{-0.5cm}
\caption{If we visualize our 3D space as a 2D surface, a wormhole can be represented as a cylindrical surface connecting two regions in the same universe or two different universes. The entrances are the mouths of the wormhole. The mouths are connected by the throat of the wormhole. \label{f-wh}}
\end{figure}

Wormholes are hypothetical spacetime structures with non-trivial topology capable of connecting either two distant regions of the same universe or two different universes, as illustrated in Fig.~\ref{f-wh}. The entrances of a wormhole are called the ``mouths'' of the wormhole and the spacetime region connecting the mouths is called the ``throat'' of the wormhole. The simplest wormhole configuration has two mouths connected by a throat, but more complex structures are also possible~\cite{Emparan:2020ldj}. Strictly speaking, wormholes are not a prediction of general relativity or of other theories of gravity. They are spacetime structures that can potentially exist in curved spacetimes, so in a very wide class of gravity models. Formation mechanisms and stability of these spacetime structures depend on the specific gravity theory and often present problems, so the existence of wormholes in the Universe is challenging. However, they cannot be ruled out.

The first wormhole solution was proposed by Einstein and Rosen in 1935 and is known as the Einstein-Rosen bridge~\cite{Einstein:1935tc}. Misner and Wheeler coined the term ``wormhole'' in Ref.~\cite{Misner:1957mt}. Among the wormhole solutions, ``traversable wormholes'' are of particular interest: they can be crossed in both directions and they could thus be potential shortcuts for interstellar or intergalactic trips to move from one region to another region of the Universe that would be unfeasible to reach within human timescales without exceeding the speed of light. Traversable wormholes were discovered in 1973 by Bronnikov~\cite{Bronnikov:1973fh} and, independently, Ellis~\cite{Ellis:1973yv}, but they became popular after the papers by Morris and Thorne~\cite{Morris:1988cz,Morris:1988tu} and Visser~\cite{Visser:1989kh}. In general relativity, traversable wormholes require exotic matter to keep the mouths open (see, however, Ref.~\cite{Bronnikov:2018uje}). In the absence of exotic matter, the wormhole throat collapses and the mouths close~\cite{Morris:1988cz,Morris:1988tu,Visser:1989kh}. The initial wormhole may become a normal black hole~\cite{Shinkai:2002gv}. However, in other theories of gravity wormholes may exist even without exotic matter or without matter at all; see, for instance, Refs.~\cite{Gravanis:2007ei,Richarte:2007zz,Eiroa:2008hv,Richarte:2010bd}.

Since we do not know if general relativity is the correct theory of gravity, nor if exotic matter can exist in the Universe, wormholes, and traversable wormholes in particular, are an exotic but fascinating idea. Astrophysical observations can look for wormholes in the Universe and in the past three decades there have been a number of proposals to search for evidence of the existence of these objects. The past 5~years have seen significant advancements of our observational facilities~\cite{Abbott:2016blz,Akiyama:2019cqa}, which has further encouraged the study of observational methods to test the existence of wormholes in the Universe.

The aim of this article is to review motivations and efforts to search for astrophysical wormholes in the Universe. Even if we will simply call them wormholes, we will only consider macroscopic traversable wormholes. The content of the manuscript is as follows. In Section~\ref{s-form}, we discuss possible formation mechanisms and stability of traversable wormholes. In Section~\ref{s-obs}, we list current attempts to find observational evidence of the existence of these spacetime structures. Summary and conclusions are in Section~\ref{s-con}.


\section{Formation and stability \label{s-form}}

The possibility of the existence of wormholes in the Universe requires that there are astrophysical mechanisms capable of creating these structures and that, once created, wormholes are stable or at least their lifetime is long enough to permit their detection. In classical general relativity, the existence of mechanisms to create traversable wormholes is challenged by some topological censorship theorems~\cite{Friedman:1993ty,Galloway:1999bp}, but quantum effects can circumvent these theorems~\cite{Gao:2016bin,Maldacena:2018lmt,Horowitz:2019hgb}.

\begin{figure}[b]
\centering
\includegraphics[width=8.5cm,trim={0cm 0cm 0cm 0cm},clip]{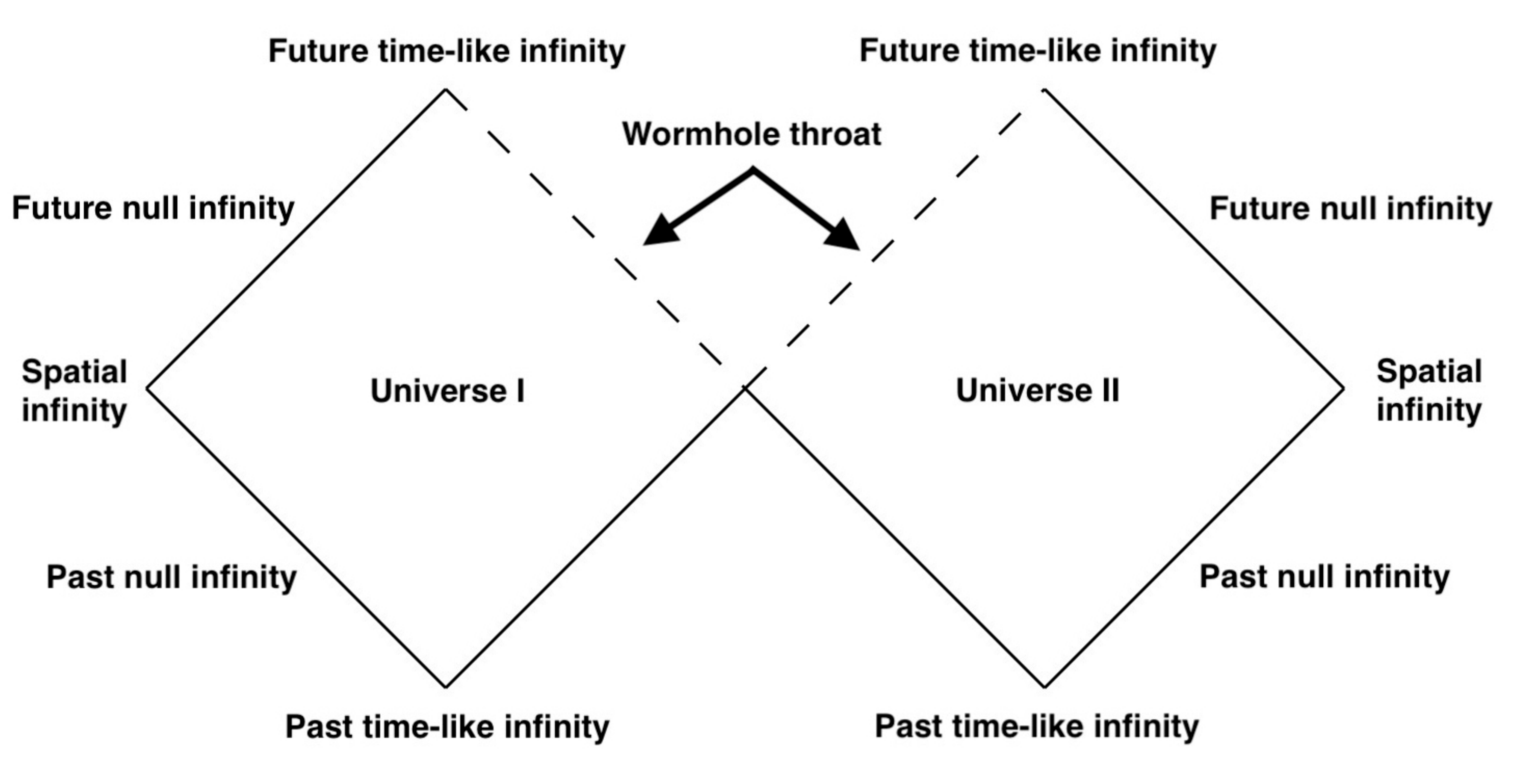}
\caption{Penrose diagram of the non-traversable Einstein-Rosen wormhole. The wormhole throat, which is a horizon, is represented by the two dashed lines: every point on the wormhole throat corresponds to the two points at the same height on the two dashed lines. \label{f-pd1}}
\end{figure}

The first wormhole solution was originally constructed by Einstein and Rosen in \cite{Einstein:1935tc}. If we start from the static black hole metric in the Schwarzschild form
\begin{equation}
ds^2=-(1-\frac{2GM}{r})dt^2+\frac{dr^2}{1-\frac{2GM}{r}}+r^2d \Omega
\end{equation}
and apply a simple coordinate transformation, $u^2=r-2 G M$, we find
\be
ds^2&=&-\frac{u^2}{u^2+2GM}dt^2+4(u^2+2GM)du^2\nonumber\\
&& +(u^2+2 G M)^2d \Omega \, .
\ee
This metric contains two asymptotically flat spacetimes, $u>0$ and $u<0$, which are connected at $u=0$. Since $u=0$ is just the $r=2GM$ hypersurface, this construct represents two disconnected spacetimes which share one horizon. 
Its Penrose diagram is shown in Fig.~\ref{f-pd1}.
Such a wormhole is non-traversable since it contains an event horizon, but it is not difficult to construct a traversable one by cutting the Schwarzschild space at $r>2GM$ and pasting the two copies together, as we will do in Section~\ref{os}.
In this representation, the wormhole connects two different universes, and as such it does not allow for shortcuts connecting separate points in the same universe.

To circumvent this last feature, these two universes are sometimes artificially connected at infinity, to make the whole construct looks as if it were one single universe. Since the geometry is static, one can argue that two distant points are in causal contact since the signal has infinite time to travel between them (outside the wormhole).  However, since the two black holes are still infinitely far away, they cannot have any useful communication or interaction outside the wormhole. This might not be a serious problem; however it renders this solution useless when such interaction is needed, for example, as in the ER=EPR conjecture \cite{Maldacena:2013xja}.

Usually, wormhole mouths located in two disconnected regions are connected by an elongated throat. The shape of the throat is given by a profile function. 
Since gravity is always attractive for positive mass distributions, the throat will almost always be unstable to small perturbations and collapse under its own gravity. 
To stabilize the wormhole, we need a negative energy distribution or some other elaborate construct 
which will provide repulsive force and keep the throat from collapsing. Since we still do not know how to create or accumulate macroscopic amounts of negative energy, it is currently not clear how we could produce a stable wormhole.  However, one of the obvious ways to minimize the amount of exotic matter is to have an ultra-short throat. 
In addition, it was argued in \cite{Visser:2003yf} that an arbitrarily thin layer of negative energy density is sufficient to keep a macroscopic wormhole stable.

So what is a possible mechanism for wormhole formation? As mentioned in the introduction, one obvious source is the quantum gravity phase of the very early
Universe. In early Universe we expect wild fluctuations in geometry and topology of spacetime to be very common. In addition, quantum effects can easily provide negative energy density which is necessary to stabilize a wormhole. Such configurations will be microscopic, but the expansion of the Universe (in particular inflation) might make them macroscopic and thus potentially observable \cite{Deng:2016vzb}. A similar mechanism has been already argued in the context of the superstring theory, where small fundamental superstrings  get stretched by inflation to render astrophysically large ``cosmic superstrings''  \cite{Polchinski:2004yav}.   

A simple method for macroscopic wormhole creation was described in \cite{Dai:2020rnc}.
There are many models in literature  in which our universe is a $3+1$-dimensional sub-space (or brane) embedded in a higher dimensional space \cite{extradim,extradim2,extradim3}. Such a brane does not have to be straight, and can bend, twist even cross itself. Therefore, some of the space points may be far apart along brane but very close in the bulk.  
A space which is folded, can potentially support a shortcut between two distant points. Alliteratively, a wormhole can connect two disconnected parallel branes. Locally, these two situations do not differ since for the local physics it is not crucial to consider how these  two branes connect in the distance.  
If we place two massive objects (e.g. black holes) in two different universes  modeled by two parallel $3+1$-dimensional branes, the exact configuration will be fully determined by the competition between two effects - the standard gravitational attraction tries to make these two objects touch, while the brane tension provides repulsion and tries to prevent this. As a result, the branes cannot stay parallel, and are  bent. If gravitational interaction is strong enough, the brane tension will not be sufficient to keep the objects apart, and they will touch. When these two branes get connected, the whole structure resembles a wormhole.  It was shown in \cite{Dai:2020rnc} that the parameter space that allows for wormhole creation is not restrictive at all. More massive and compact objects are more likely to fulfill the conditions for such wormhole-like formation, which
implies that we should be looking for realistic wormholes either in the background of black holes and compact stars, or massive microscopic relics. In this construct, the role of negative
energy density which provides repulsion that counteracts gravity is played by the brane tension. Thus we
do not need extra source of negative energy density to support gravity. However, this still does
not guaranty stability of the whole construct. It could happen that a very long wormhole throat breaks into
smaller pieces in order to minimize its energy. 

Another way to balance gravity is to place a wormhole in a de Sitter background. In \cite{Dai:2018vrw}
 a wormhole solution was constructed by placing two black holes at the
antipodes of the closed de Sitter space. The gravitational action of the  cosmological constant provides the repulsive force and counteracts attractive gravity between the black holes,
and makes the whole configuration static. 


\section{Search for astrophysical wormholes \label{s-obs}}

Most wormholes studied in the literature are black hole mimickers. In such a case, we can employ electromagnetic~\cite{Bambi:2015kza} and gravitational wave techniques~\cite{Yagi:2016jml} that have been proposed to test the nature of astrophysical black holes. These methods normally test the spacetime metric around the source, whether it is the Kerr solution of uncharged black holes in general relativity or another metric, and are not designed to distinguish generic black holes from generic wormholes. However, there are also some methods to distinguish generic black holes from generic wormholes, which are based on the idea that we can somehow observe bodies in the spacetime region on the other side of the wormhole throat. Note also that some wormholes are not black hole mimickers; for example, they may not have a (effective positive) mass and therefore they cannot have orbiting or accreting material around them. However, these wormholes can still have an impact on the motion of passing particles (e.g., deflection of light in micro-lensing events).

\subsection{Gravitational lensing}

\begin{figure*}[t]
\centering
\includegraphics[width=7.5 cm,trim={3.5cm 14cm 1.5cm 2.5cm},clip]{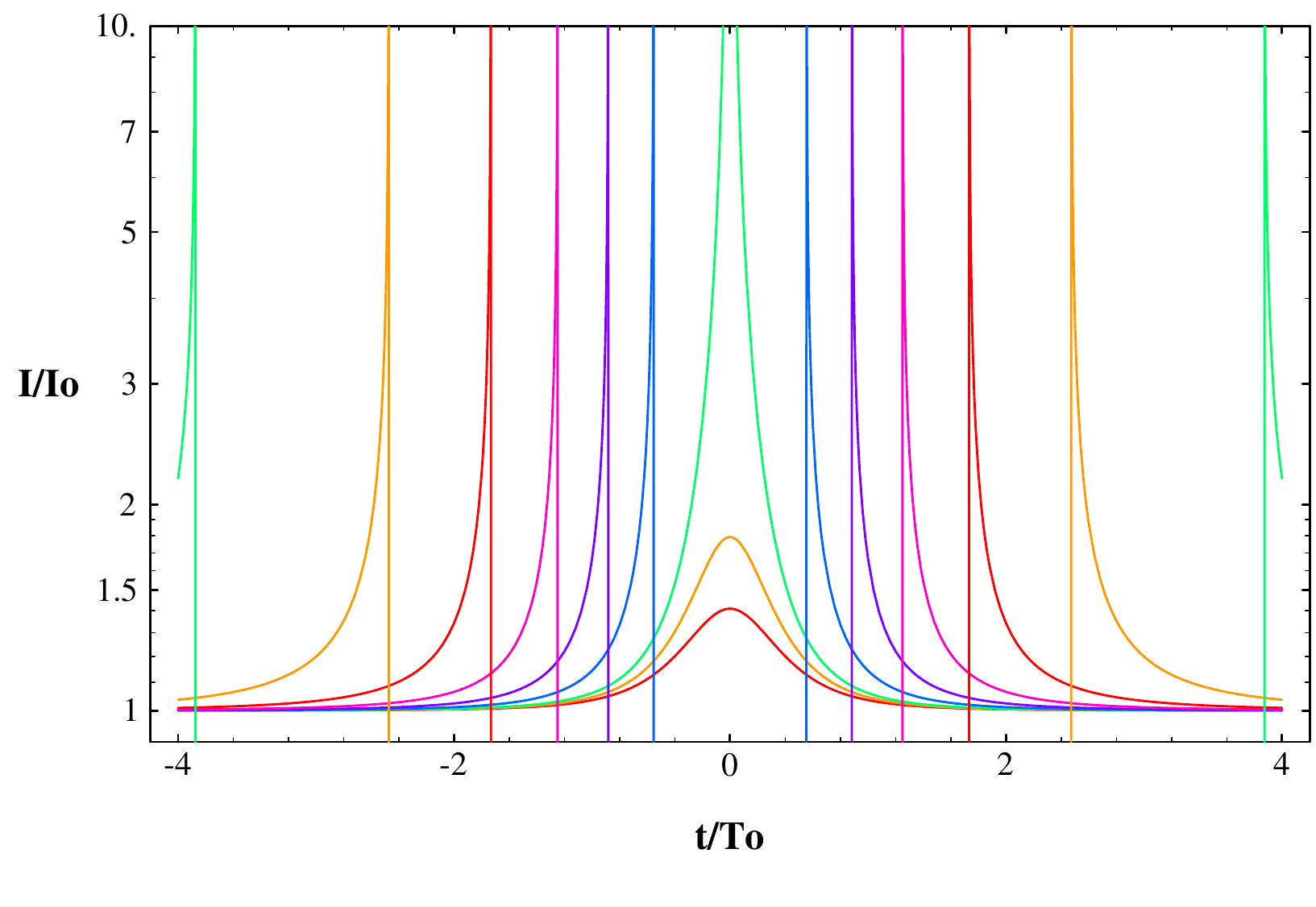}
\hspace{1.0cm}
\includegraphics[width=7.5 cm,trim={3.5cm 14cm 1.5cm 2.5cm},clip]{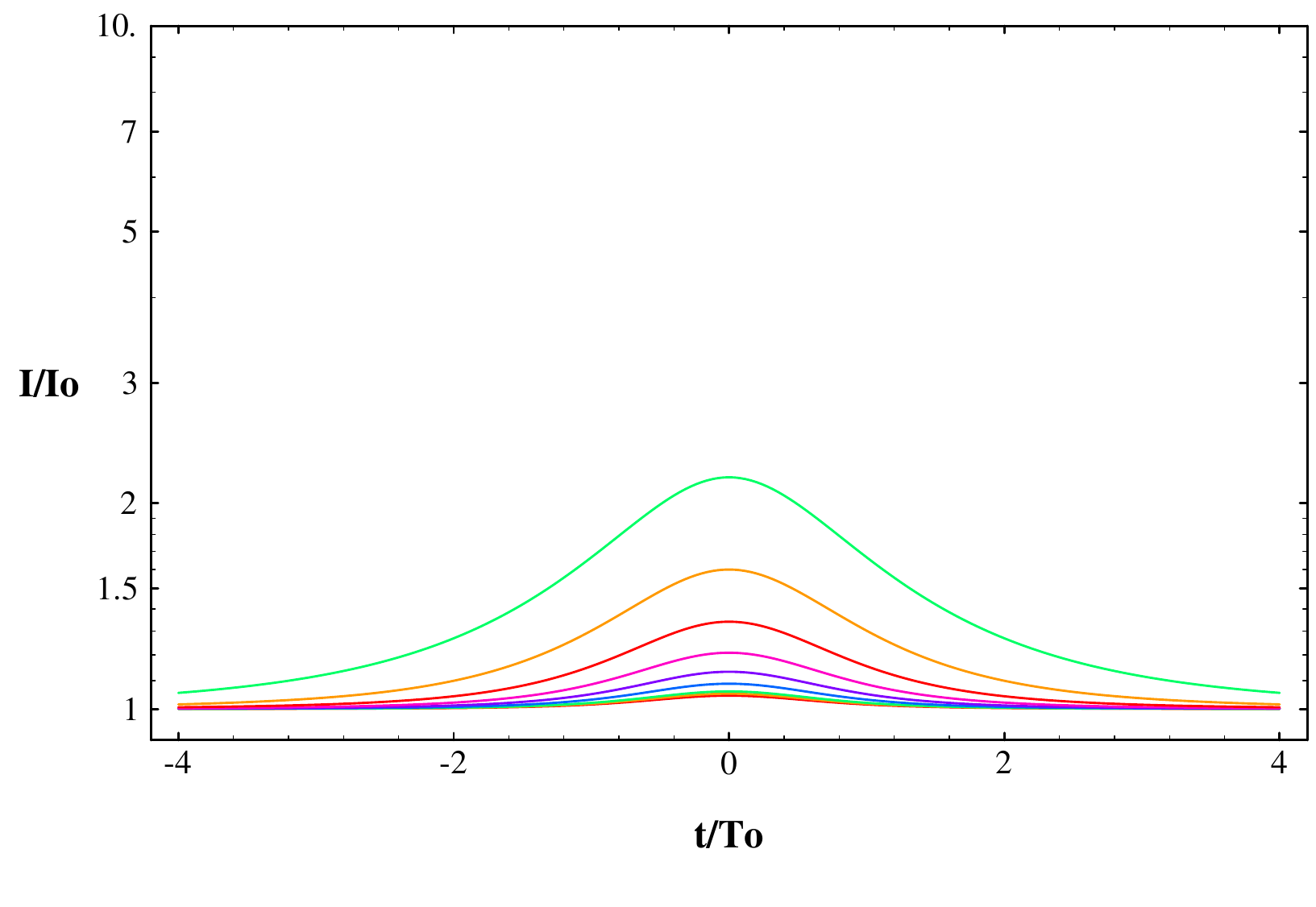}
\vspace{-0.4cm}
\caption{Evolution of the intensity of a background source at the passage of a wormhole with effective negative mass (left panel) and a normal compact object (right panel). Different curves correspond to different values of the impact parameters (from the smallest value of the impact parameter producing larger effects to the largest value of the impact parameter producing less pronounced modulation of the intensity of the source: green-orange-red-magenta-violet-blue-green-orange-red). Figure from Ref.~\cite{Cramer:1994qj}. \label{f-ml}}
\end{figure*}

The idea to look for astrophysical wormholes by searching for specific micro-lensing events was historically the first strategy to try seriously to find evidence for the existence, or at least to constrain the abundance, of wormholes~\cite{Kim:1993ma,Cramer:1994qj}. Fig.~\ref{f-ml} shows the evolution of the intensity of a background source at the passage of certain wormholes with effective negative mass (left panel) and of normal compact objects (right panel). In Ref.~\cite{Torres:1998xd}, it is shown that certain wormholes can produce micro-lensing events that can resemble certain gamma-ray bursts and, using \textsl{BATSE} data, the authors infer an upper limit to the mass density of wormhole-like objects in the Universe.

\begin{figure}[t]
\centering
\includegraphics[width=7.0 cm]{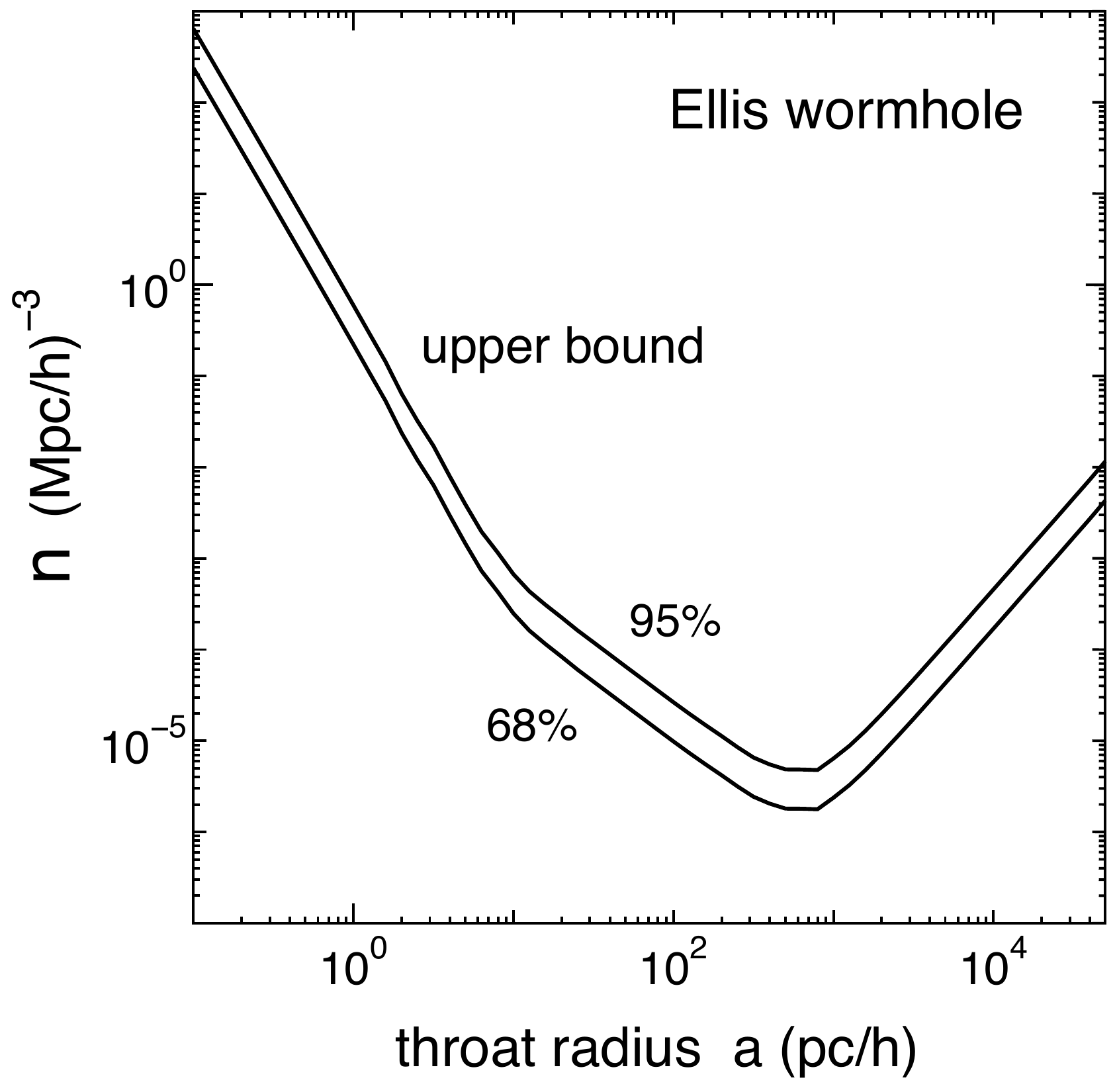}
\vspace{-0.2cm}
\caption{Upper bounds at the 68\% and 95\% confidence levels on the number density of Ellis wormholes in the Universe from the Sloan Digital Sky Survey Quasar Lens Search. The horizontal axis is for the throat radius $a$ (in units pc/$h$, where $h$ is the scaling factor for the present-day Hubble expansion rate). Figure from Ref.~\cite{Takahashi:2013jqa}. \label{f-ellis}}
\end{figure}

The gravitational lensing of a number of wormhole solutions has been studied; see, e.g., Refs.~\cite{Torres:2001gb,Nandi:2006ds,Rahaman:2007am,Dey:2008kn,Bhattacharya:2010zzb,Nakajima:2012pu,Tsukamoto:2012xs,Kuhfittig:2013hva,Tsukamoto:2016qro,Shaikh:2017zfl,Jusufi:2017vta,Ovgun:2018fnk,Ovgun:2018prw,Ono:2018ybw,Shaikh:2018oul,Shaikh:2019jfr}. Observational bounds on the possible abundance of these structures are reported in Refs.~\cite{Anchordoqui:1999gca,Abe:2010ap,Toki:2011zu,Takahashi:2013jqa}. Fig.~\ref{f-ellis} shows the observational bounds from the Sloan Digital Sky Survey Quasar Lens Search on the number density of Ellis wormholes as a function of the throat radius $a$ as inferred in Ref.~\cite{Takahashi:2013jqa}.

We note that exotic structures like wormholes may even cause demagnification (instead of magnification) of background sources~\cite{Kitamura:2012zy}, and such a phenomenon can be used to search and constrain some kinds of wormholes (and other exotic objects) in gravitational lensing surveys. There are also attempts to search for wormholes by the studying the shear~\cite{Izumi:2013tya} and the light time delay~\cite{Nakajima:2014nba} of gravitational lensing events.

\subsection{Orbiting stars} \label{os}

If a traversable wormhole smoothly connects two different spacetimes, then fluxes (including gravitational flux) cannot be
separately conserved in any of these spaces individually. Then objects propagating in a vicinity of a
wormhole in one space must feel influence of objects propagating in the other space. This leaves us with very interesting possibility of observing a wormhole by studying the orbits of stars on our side of a potential wormhole. 

\begin{figure}[t]
\centering
\includegraphics[width=7.5cm,trim={0cm 0cm 0cm 0cm},clip]{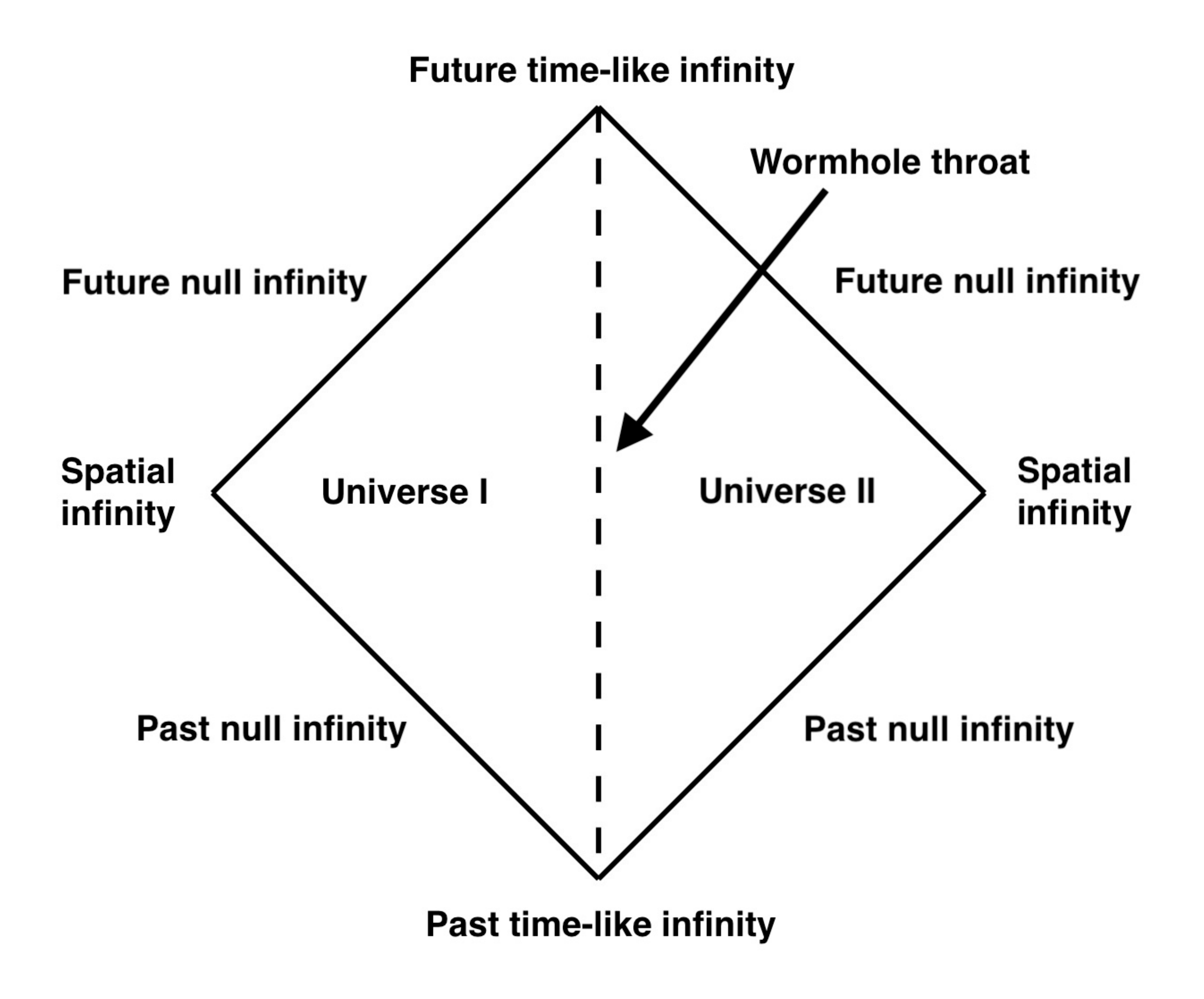}
\vspace{-0.3cm}
\caption{Penrose diagram of the traversable wormhole discussed in Section~\ref{os}. The wormhole throat is the horizontal dashed line and connects the two universes. \label{f-pd2}}
\end{figure}

To model a traversable wormhole harbored by a massive compact object, we cut the Schwarzschild spacetime at the radius $R$ which is slightly larger than the gravitational radius, i.e. $R \ge r_g \equiv 2GM$. We take another identical spacetime and paste them together. Our global construct is thus  two copies of the Schwarzschild spacetime connected through a mouth of radius $R$. 
Its Penrose diagram is shown in Fig.~\ref{f-pd2}.
Some exotic matter with negative energy density is needed to keep the wormhole stable; however, in the short throat approximation that we use, we assume that the effects of this exotic matter are subdominant.  We now place a perturber on the other side of the wormhole and a test object (or an observer) on our side. By deriving the gravitational perturbations on this background, we can show \cite{Dai:2019mse,Dai:2019nph} that  an observer on our side will feel an additional (anomalous) acceleration due to perturbations sourced on the other side
\begin{equation}
\label{acceleration}
a\approx-\mu \frac{R}{A}\frac{1}{r^2} \, ,
\end{equation}
where 
$\mu$ is the mass of the perturber on the other side, $A$ is the radial location of the perturber, and $r$ is the radial distance from the center of the wormhole on our side.
If all we had was just this monopole contribution, it would be very difficult to extract an observable effect, since this additional acceleration would just simply add to the acceleration due to the central object. We therefore consider an elliptic orbit of a perturber, i.e. an object orbiting on the other side of the wormhole with the periapsis radius $r_p$ and apoapsis radius $r_a$. An elliptic orbit cannot be represented with only one monopole, and can be viewed instead as a sequence of monopoles. We estimate the magnitude of the acceleration variation by using two monopoles, one for a pericenter, $r_p$, and another for an apocenter, $r_a$, as 
\begin{equation}
\Delta a =\mu R \left(\frac{1}{r_p}-\frac{1}{r_a}\right)\frac{1}{r^2} \, .
\end{equation}
If the orbit of an object on the other side of the wormhole's is elongated so that $r_a\gg r_p$, then we can approximate the magnitude as
\begin{equation}\label{da}
\Delta a =\mu \frac{R}{r_p}\frac{1}{r^2} \, .
\end{equation}

Note that what we calculate in our Eq.~(\ref{da}) is the magnitude of acceleration variation of an object in our space due to an elliptic orbit of a perturber on the other side perturbing the metric. These variations come on top of the constant acceleration that
comes from the central object. With good enough precision, we should be able to detect or exclude this variable anomalous acceleration. Of course, these variations can possibly be produced by some other dim sources on our side. Then, more careful modeling would be required to distinguish between different options.

\begin{figure}
   \centering
\includegraphics[width=6cm]{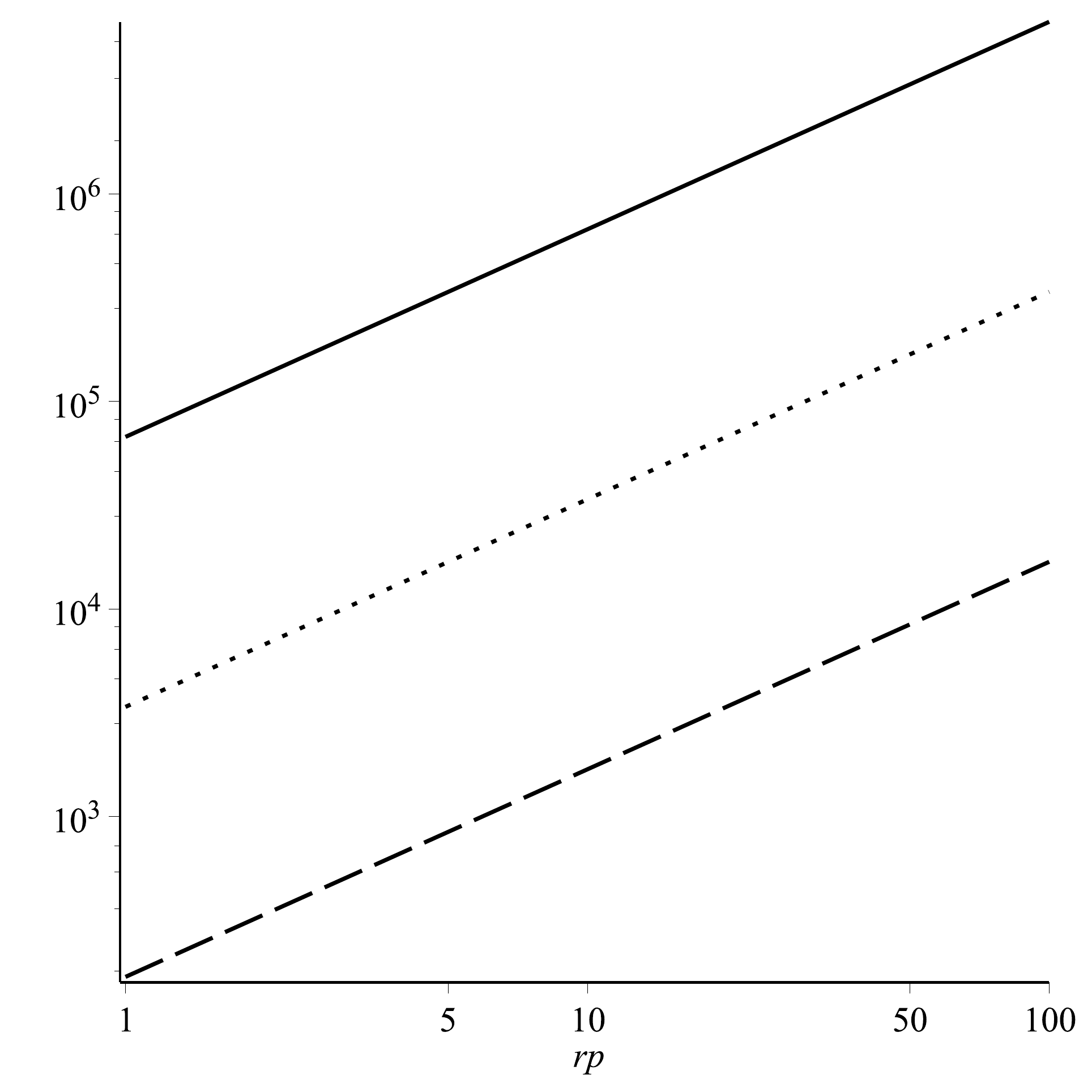}
\vspace{-0.2cm}
\caption{We plot the constraints on the mass $\mu$ (y-axis) and the periapsis radius $r_p$ (x-axis) of a hypothetical star that orbits Sgr A* on the other side and perturbs the orbit of the S2 star on our side.
The solid, doted, and dashed lines represent the constraints with acceleration precision of the star S2 of $4\times 10^{-4} m/s^2$, $2\times 10^{-5} m/s^2$, and $10^{-6} m/s^2$ respectively. The regions above the  lines rule out a wormhole explanation. The x-axis has units of $r_g=2GM$. The y-axis has units of the solar mass $M_\odot$. The bottom line probes the most reasonable parameter space - a few solar masses star orbiting around Sgr A* at the distance of a few gravitational radii.
}
\label{constraint}
\end{figure}

We can then apply Eq.~(\ref{da}) to a concrete example  - the star S2 orbiting the center of our Galaxy Sgr A*. We believe that  Sgr A*
contains a supermassive black hole with $M = 4 \times 10^6 M_\odot$. The star S2, which orbits Sgr A* has a mass of $14 M_\odot$, with an orbital period of 15.9 years, and a semi-major axis of 1031.69 AU. So far we measured the total acceleration of the star S2 in orbit to be $1.5m/s^2$, which mainly comes from the supermassive black hole. This
acceleration has been measured with the precision of $4\times 10^{-4} m/s^2$. Within a decade this can be improved to $10^{-6}m/s^2$. As Fig.~\ref{constraint}  shows, with this precision one can exclude the most generic situation - a few solar masses star orbiting around Sgr A* on the other side of the wormhole at the distance of a few gravitational radii. If we do not notice any detectable imprint on the orbit of the S2 star with this precision, a wormhole interpretation of Sgr A* will perhaps be eliminated. To test the wormhole hypothesis, one would need to exclude all the other more conservative explanations, like perturbers on our side or higher order relativistic effects~\cite{Do:2019txf}.

The most direct way to observe the effect of the anomalous acceleration is to look for deviations of the object's orbit from the expected, unperturbed Keplerian or general relativistic result. In particular, we can look for additional periodic variations in the orbital velocities of objects on our side. A detailed analysis of this effect was performed in \cite{Simonetti:2020vhw}.
To estimate the change in the orbital velocity caused by $\Delta a$ given in Eq.~(\ref{acceleration}) we can assume, for simplicity, that the additional acceleration occurs once every orbital period $T\,'$ of the perturber (i.e., when it is near its periapsis).

We consider systems where the duration of the additional acceleration $\Delta a$ (i.e., the time the perturber is near its periapsis) is much shorter than the orbital period of the perturbed star on our side of the wormhole, $T$, so we treat the effect of the perturber on the object we observe as impulsive. 
After modeling and removing the unperturbed orbital behavior of the Doppler velocity of our observed object, if an additional cyclic variation of some period $T\,'\ll T$ is not readily apparent, the best strategy to search for such a result is to cut the sequence of velocity residuals into segments of some duration $T\,'$. Then stack and average the sequences. In this way one could detect a cyclic variation in Doppler velocity of period $T\,'$ as the noise in the resulting measurements is reduced by $\sqrt{\tau/T\,'}$, where $\tau$ is the duration of the observing program. Comparing the measurement error with variation coming from an anomalous acceleration, we can get an upper mass limit on the perturber as
    \begin{equation}\label{masslimit}
    \mu > \frac{1}{G} \frac{r_p'}{r_g}\ r_{avg}^2\ \frac{1}{f'^2 T\,'} \ \sigma_v 
    \left(  \frac{T\,'}{\tau} \right)^{1/2}
\end{equation}
where $r_{avg}$ is the average distance of the observed object, on our side, from the wormhole (i.e., the semi-major axis of the object's orbit), $f'=r_p'/r_a'$, while $\sigma_v$ is the measurement uncertainty in velocity. Since $T\,'\propto {r_p'}^{3/2}$ from the Kepler's laws, this limit is $\propto {r_p'}^{1/4}$.

For the generic values of the parameters we can see that a better limit could be set from observations of a star in orbit around a stellar-mass black hole, instead of the supermassive black hole as Sgr A*. The cleanest systems would be those with no mass transfer, which would avoid dynamical changes not caused by a perturber on the other side of the wormhole. For example, for 10$M_\odot$ black hole, stellar orbit of radius $\sim10^6 r_g$, and individual Doppler velocity measurements for the star with $\sigma_v \sim 6$~km/s, 
the perturber mass limit for this case could be $\sim$4 orders of magnitude lower than obtained from observations of S2 for $\tau =1$~year.

Observations of a pulsar orbiting a black hole have the potential to set even better limits, given the greater observational precision attainable. Black hole-pulsar binaries have been argued to provide remarkable tests of quantum gravity \cite{mike1,mike2,mike3,mike4,mike5,mike6,mike7} on top of their proven record in testing Einstein's general relativity in the case of the Hulse-Taylor binary pulsar PSR~1913+16 \cite{weisberg}.

 The precision on measured parameters for a pulsar is determined by the precision on pulse ``times of arrival'' (TOA) measurements, which can be as low as  $\sigma_{TOA}\sim 0.1\mu$s. 
For a pulsar in an orbit around Sgr A* which is similar to that of S2, a mass limit for the perturber could be $\sim 4$ orders of magnitude lower than for observations of S2.
Still better results could be obtained for pulsars in close orbits around stellar-mass black holes. Consider the ``nominal'' case of a pulsar in orbit around a $10M_\odot$ black hole where $r_{avg}\approx2\times10^9$~m, the semi-major axis for the Hulse-Taylor pulsar. For observations over $\tau=1$~year, and $\sigma_v\sim$1~m/s we obtain a limit on the perturber mass more than $\sim6$ orders of magnitude better than for a pulsar orbiting Sgr~A* (see Ref.~\cite{Simonetti:2020vhw}).

\subsection{Imaging}

Efforts for imaging supermassive black holes and observing the so-called black hole ``shadow'' started more than 20~years ago~\cite{Falcke:1999pj}. The boundary of the shadow of a black hole corresponds to the photon capture sphere as seen by a distant observer and its analysis can thus constrain the spacetime geometry around the compact object~\cite{Bambi:2008jg}. The possibility of testing the nature of supermassive black holes from the study of their shadow has become quite a hot topic after the release of the image of the object at the center of the galaxy M87 by the Event Horizon Telescope Collaboration~\cite{Akiyama:2019cqa}.

\begin{figure*}[t]
\centering
\includegraphics[width=6.0cm]{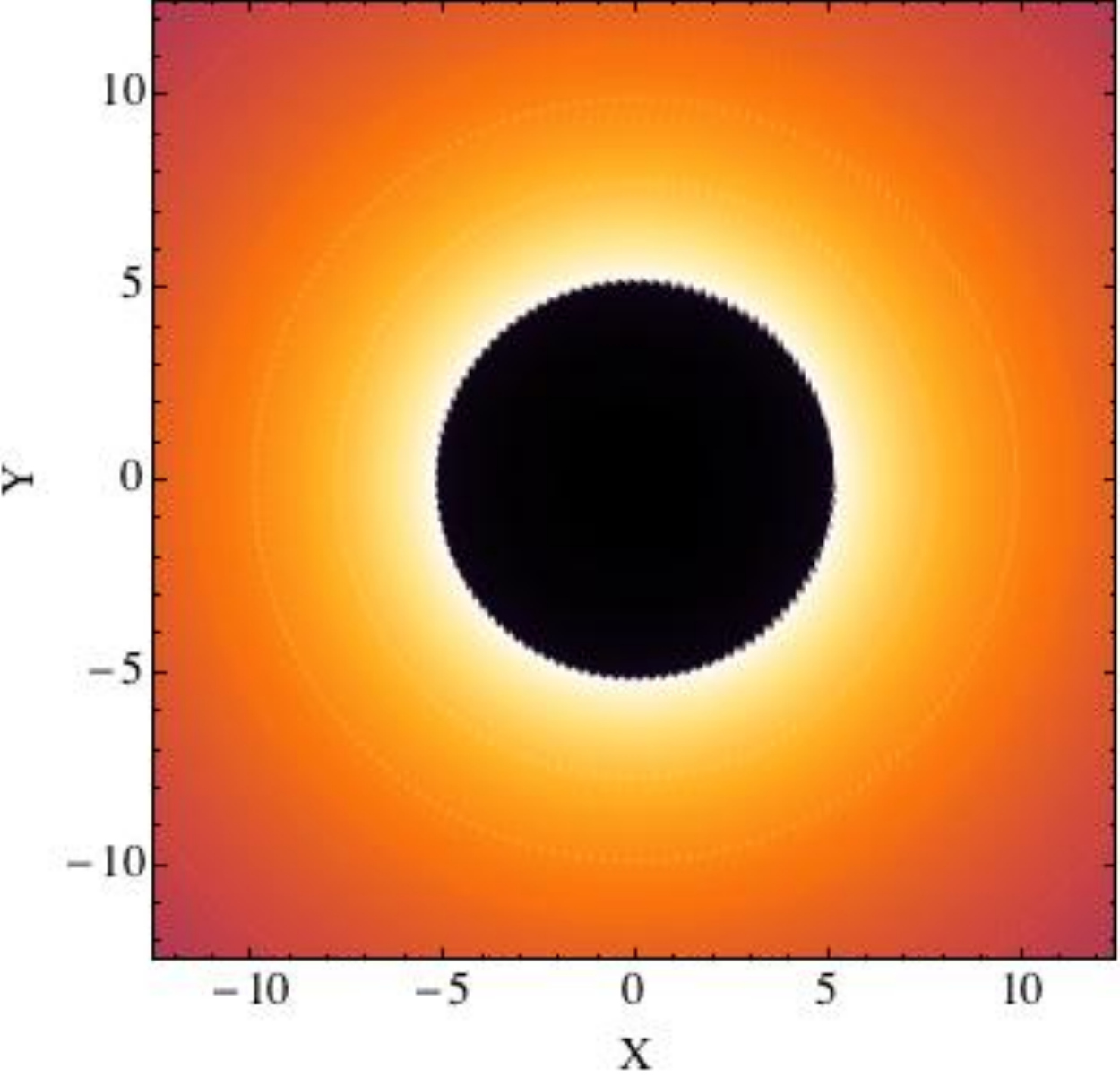}
\hspace{1.0cm}
\includegraphics[width=6.0cm]{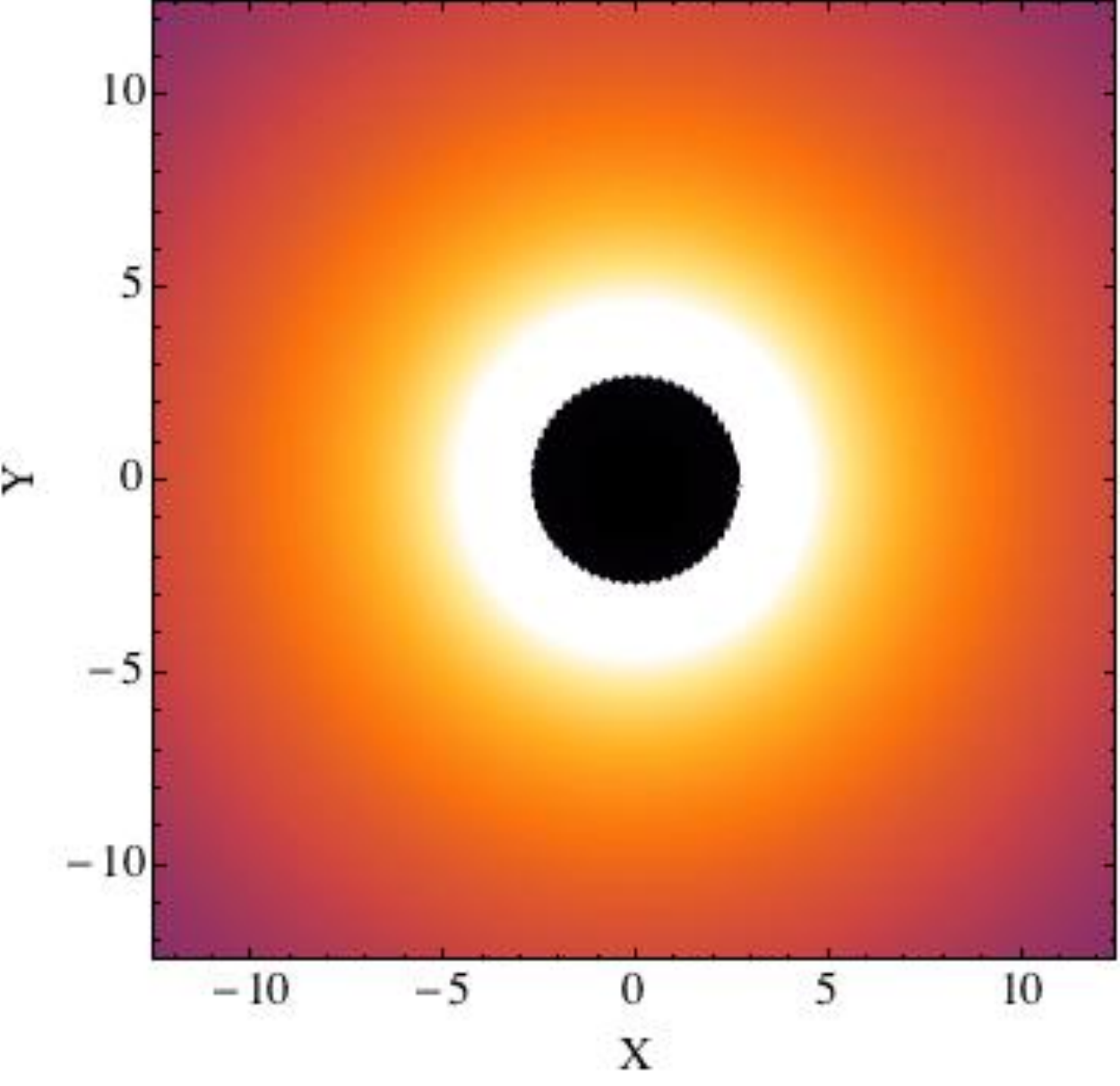}
\caption{Simulated image of an optically thin emission region surrounding a Schwarzschild black hole (left panel) and a traversable spherically-symmetric wormhole (right panel). The coordinates $X$ and $Y$ are in units of the gravitational radius of the system. Figure from Ref.~\cite{Bambi:2013nla}. \label{f-sh}}
\end{figure*}

Most wormhole spacetimes have a photon capture sphere outside of their mouths. If a wormhole is surrounded by some optically thin emitting gas, a distant observer can thus see the shadow of the wormhole. The first study on a wormhole shadow was reported in Ref.~\cite{Bambi:2013nla}, where it was pointed out that the size of the shadow of a wormhole can be quite different from that of a black hole because of the very different light bending, see Fig.~\ref{f-sh}. This is different from the case of black holes of different gravity theories that normally have quite similar shadows and it is thus very challenging to test different black hole solutions~\cite{Mizuno:2018lxz}. The reason can be understood as follows. Traversable wormholes do not have a horizon because the gravitational field around these objects is weaker than the gravitational field around black holes, so we can expect that even their photon capture sphere and shadow are normally smaller than those of black holes. The wormhole solution of Ref.~\cite{Bambi:2013nla} can already be ruled out by the current image of the object at the center of the galaxy M87~\cite{Akiyama:2019cqa}.

A number of studies on the shadows of wormholes are now present in the literature~\cite{Nedkova:2013msa,Ohgami:2015nra,Abdujabbarov:2016efm,Shaikh:2018kfv,Gyulchev:2018fmd,Amir:2018szm,Wang:2020emr,Wielgus:2020uqz}. Every wormhole type has its own shadow. In the case of spherically symmetric wormholes, the boundary of the shadow is a circle and the wormhole scenario can only be tested if the shadow size is very different from that expected for black holes (or if we have independent estimate of the spin and the inclination angle of the source); indeed even the shadow of a Schwarzschild black hole and the shadow of a Kerr black hole observed along its spin axis are circles. In the case of axisymmetric wormholes, the boundary of the shadow is a ``deformed'' circle and, depending on the specific model, may be tested if we are not observing the source along its axis of symmetry.

A test similar to the analysis of the shadow is represented by the possibility of imaging a blob of plasma orbiting a supermassive object~\cite{Li:2014coa}: even in this case, the apparent image of the blob of plasma would be quite different because of the difference in the light bending of wormhole and black hole spacetimes.

An alternative test, which is based on the intrinsic properties of traversable wormholes rather than their model-dependent spacetime metric, is represented by the possibility of detecting radiation emitted from the other side of the wormhole~\cite{Shatskiy:2008ym,Doroshkevich:2008xm}. Such a test would have really the power to distinguish a wormhole from a black hole, even if the opposite would not be true. Note, however, that current imaging techniques are at mm wavelength, so we cannot observe the optical image of a star on the other side of the throat of the wormhole.

\subsection{Accretion disk spectra}

The analysis of the properties of the radiation emitted from the inner part of the accretion disk around a compact object can be a powerful tool to probe the geometry in the strong gravity region and thus test the nature of the source~\cite{Bambi:2015kza}. The best systems for these tests are those with a geometrically thin and optically thick accretion disks, which requires that the accreting material has a large angular momentum and that the Eddington-scaled mass accretion rate of the source is around 10\%.

\begin{figure}[t]
\centering
\includegraphics[width=8.5 cm]{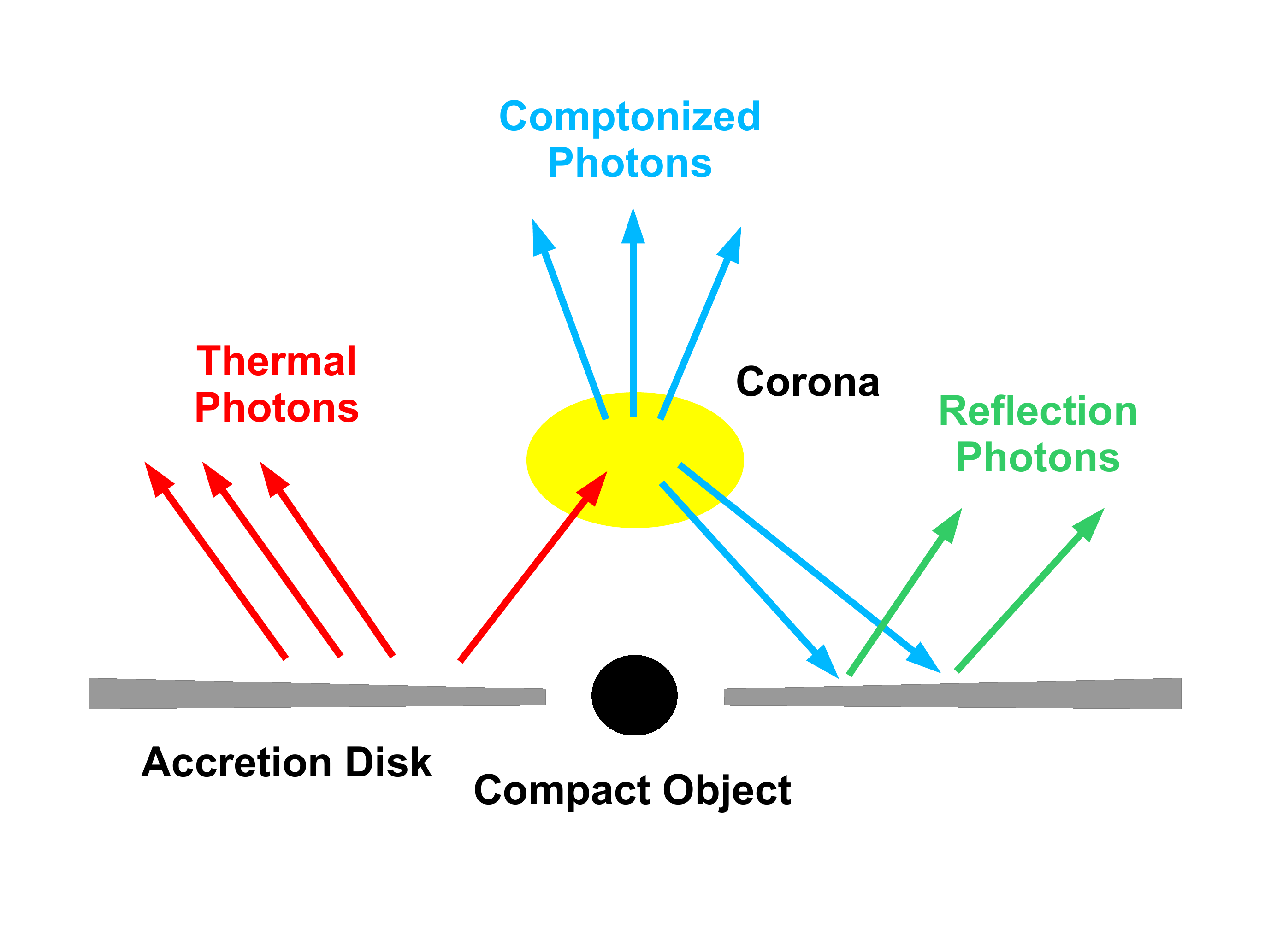}
\vspace{-0.8cm}
\caption{A compact object accreting from a geometrically thin and optically thick disk. We highlight the multi-temperature blackbody spectrum of the disk (red), the Comptonized photons (blue), and the reflection spectrum (green). Figure adapted from Ref.~\cite{Bambi:2017iyh}. \label{f-disk}}
\end{figure}

The system is illustrated in Fig.~\ref{f-disk}. The compact object is represented by the black circle and can be either a wormhole or a black hole. The standard framework for the description of geometrically thin and optically thick accretion disks is the Novikov-Thorne model~\cite{Novikov:1973kta,Page:1974he,Bambi:2017khi}, which only requires that the spacetime is stationary, axisymmetric, and asymptotically flat. From the conservation of mass, energy, and angular momentum, we can infer the radial structure of the disk. Every point on the disk is in local thermal equilibrium and has a blackbody-like spectrum. The whole disk has a multi-temperature blackbody spectrum. The temperature of the inner part of the accretion disk roughly scales as $M^{-1/4}$, so the spectrum is peaked in the soft X-ray band (0.1-1~keV) for stellar-mass compact objects ($M \sim 10~M_\odot$) and in the UV band (1-100~eV) for supermassive compact objects ($M \sim 10^5$-$10^{10}~M_\odot$). Thermal photons from the disk can inverse Compton scatter off free electrons in the so-called ``corona'', which is some hotter ($T_{\rm e} \sim 100$~keV) electron cloud in the strong gravity region of the compact object. The corona may be the atmosphere above the accretion disk, the accretion flow between the inner edge of the disk and the compact object, the base of the jet, etc.~\cite{Bambi:2020jpe}. More than one corona may coexist at the same time. The Comptonized photons have a spectrum that can be approximated by a power-law (with a photon index in the range 1-3) with an exponential high-energy cutoff (of the order of the coronal temperature). A fraction of the Comptonized photons can illuminate the accretion disk: Compton scattering and absorption followed by fluorescent emission generate the reflection component. More details can be found in \cite{Bambi:2020jpe} and references therein.

In principle, the analysis of the thermal spectrum of the disk can be used to constrain the spacetime metric around a compact object and thus test the wormhole scenario. Studies of specific wormhole solutions are presented in Refs.~\cite{Harko:2008vy,Harko:2009xf,Izmailov:2019hmf,Paul:2019trt}. However, there are two facts to take into account and that limit the use of this technique to search for astrophysical wormholes. First, the method only works for stellar-mass compact objects. Indeed, for the supermassive ones the emission is peaked in the UV, where dust absorption strongly limits our capability of accurate measurements of the spectrum of the disk. Second, the thermal spectrum of the disk has a very simple shape, with the result that when we analyze real data we find a strong parameter degeneracy, which challenges the possibility of distinguishing a Kerr black hole from another compact object~\cite{Tripathi:2020qco}.

A more powerful method is represented by the analysis of the reflection features. The intrinsically narrow iron K$\alpha$ line is particularly suitable to measure the effects of relativistic blurring and thus constrain the spacetime metric of the source. This is today a mature technique that, for specific sources and observations, has shown to be able to provide precise and accurate tests of the Kerr metric~\cite{Tripathi:2020dni,Tripathi:2020yts}. It can be used to test either stellar mass and supermassive objects, as the reflection spectrum does not directly depend on the mass of the source. Preliminary studies with this technique have only considered the iron K$\alpha$ line, limiting the analysis at a qualitative level~\cite{Bambi:2013jda,Zhou:2016koy}. With the recent developments of reflection models suitable for generic stationary and axisymmetric spacetimes~\cite{Bambi:2016sac,Abdikamalov:2019yrr}, it is now possible to test wormhole scenarios with the state of the art of reflection modeling and high-quality data of current X-ray missions~\cite{Tripathi:2019trz}. As in other cases, here we test the Kerr metric against some specific wormhole solution. Some wormholes metric can already be ruled out, while in other cases the reflection spectrum is not sufficiently different from that of Kerr black holes to distinguish the two scenarios with the available data.

In the case of traversable wormholes, we might expect that some material can flow from one of the mouths to the other one, which is not the case for black holes. This process could generate $\gamma$ radiation resulting from the collision of accretion flows moving to opposite directions~\cite{Piotrovich:2020kae}. The possible detection of such a component in the spectrum of a source would be a specific signature of the presence of a throat, and thus a model-independent test to distinguish wormholes from black holes. It requires a good knowledge of the high-energy spectrum of a source, but it is doable.

\subsection{Gravitational waves}

Gravitational wave tests represent a complementary approach to search for astrophysical wormholes in the sky. Unlike electromagnetic tests, which only depend the spacetime metric (assuming that all particles follow the geodesics of the spacetime), gravitational wave tests require to calculate the gravitational wave signal from the field equations of the theory, so it is necessary to specify the theoretical framework. Alternatively, we can make the assumption that the emission of gravitational waves in a coalescing binary system can be described well by the Einstein Equations, and in such a case we can try to test the wormhole metric of gravity models beyond general relativity.

\begin{figure}[t]
\centering
\includegraphics[width=8.0 cm]{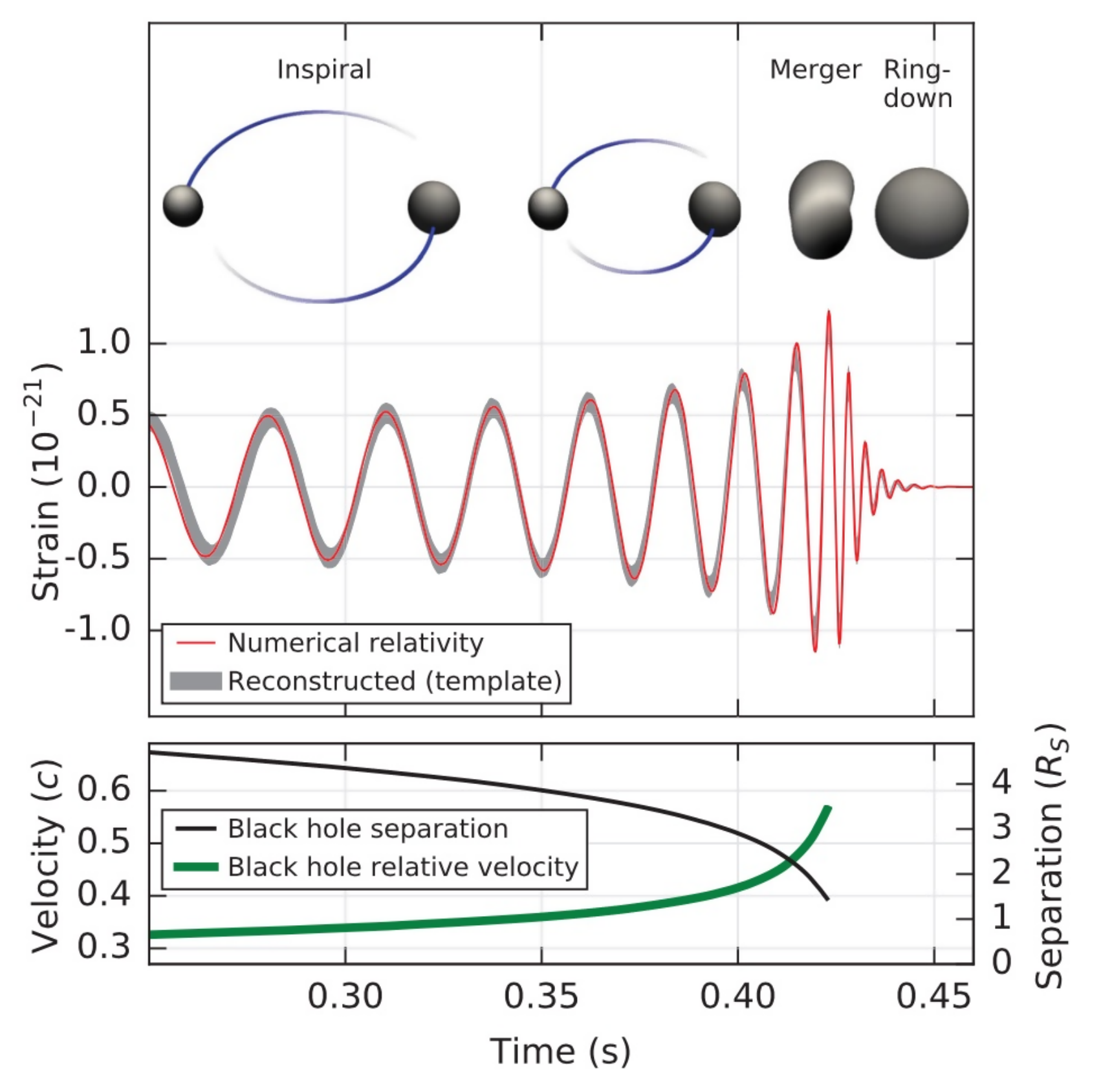}
\caption{Temporal evolution of the strain, the black hole separation, and the black hole relative velocity in the event GW150914. From Ref.~\cite{Abbott:2016blz} under the terms of the Creative Commons Attribution 3.0 License. \label{f-gw}}
\end{figure}

The LIGO and Virgo experiments have now reached the sensitivity to detect one coalescence of two stellar-mass compact objects every few days. The system is shown in Fig.~\ref{f-gw}. The event is characterized by three different stages: inspiral, merger, and ring-down. In the inspiral phase, the two bodies orbit around a common center of mass, the system loses energy and angular momentum through the emission of gravitational waves, so the separation between the two bodies decreases and the relative velocity and the orbital frequencies increase. The inspiral phase is followed by the merger, when we have the formation of one object. After that, there is the ring-down phase: the newly born object emits gravitational waves to settle down to an equilibrium configuration.

During the inspiral phase, the two objects are relatively far each other, with the exception of the time before the merger, which may be too short to make accurate and precise measurements of the spacetime geometry. As a result, it is challenging to test the geometry of the spacetime around the sources. Better measurements are possible if the mass difference between the two object is significant, and this is the case discussed in Ref.~\cite{Dent:2020nfa}, where the authors consider a $5~M_\odot$ black hole orbiting a $200~M_\odot$ wormhole. A similar unequal-mass system has not yet been observed by the LIGO/Virgo Collaboration, but it could be observed in the future. Better chances will be possible with space-based gravitational wave antennas (after 2030) as it is one of their main targets the observation of gravitational waves from stellar-mass compact objects orbiting $\sim 10^6~M_\odot$ compact objects. Wormholes can also have very different tidal properties and, if very compact, may even give rise to intermittent bursts following ring-down; see, e.g., \cite{Cardoso:2019rvt} and references therein.

The merger phase is potentially quite informative about the nature of the compact object and the gravity theory, but it is too short for accurate and precise measurements of the signal and accurate theoretical predictions are still challenging.

In the end, it seems that the ring-down phase, and especially the late-time ring-down phase, is the most suitable to test fundamental physics~\cite{Cardoso:2016rao}. During the ring-down, the newly born object emits gravitational waves to settle down to an equilibrium configuration. The emission is characterized by the so-called quasi-normal modes (QNMs), which are characteristic of the nature of the final product of the coalescence. QNMs of different wormhole solutions have thus been calculated as a tool for testing specific wormhole models with gravitational wave data~\cite{Konoplya:2005et,Konoplya:2016hmd,Nandi:2016uzg,Bueno:2017hyj,Aneesh:2018hlp,Blazquez-Salcedo:2018ipc,Churilova:2019qph}. The current gravitational wave data cannot test these models, because it is necessary to have precise and accurate measurements of at least two QNMs, but this can be achieved in the near future.


\section{Concluding remarks \label{s-con}}

The existence of traversable wormholes in the Universe is an exotic but fascinating hypothesis that cannot be ruled out at the moment. They might represent our only possibility for interstellar and intergalactic trips in a far future. Most wormhole solutions are black hole mimickers and the remarkable progress in the past few years in our capabilities of probing the strong gravity region of black holes has encouraged new studies to check whether astrophysical black holes, or at least some of them, are actually the mouths of wormholes towards faraway regions or even other universes. In this review articles, we have tried to summarize the state of the art in the search for wormholes in the Universe. Unfortunately, it has not been possible to cite and discuss all contributions in the field. However, we hope that this review can serve as a good starting point for the interested reader to explore this fascinating line of research.


\vspace{0.5cm}

{\bf Acknowledgments --}
The work of C.B. is supported by the Innovation Program of the Shanghai Municipal Education Commission, Grant No.~2019-01-07-00-07-E00035, the National Natural Science Foundation of China (NSFC), Grant No.~11973019, and Fudan University, Grant No.~JIH1512604. D.S. is partially supported by the US National Science Foundation, under Grant No.~PHY-2014021.


\end{document}